\documentstyle[aps]{revtex}
\begin{document}
\draft
\twocolumn[\hsize\textwidth\columnwidth\hsize\csname@twocolumnfalse%
\endcsname

\title{\bf Dynamics and Instabilities of Planar Tensile Cracks in
Heterogeneous Media} \author{Sharad Ramanathan and Daniel S. Fisher}
\address{Physics Department, Harvard University, Cambridge,
Massachusetts 02138} \date{\today}

\maketitle 
\begin{abstract} 

The dynamics of tensile crack fronts restricted to advance in a plane
are studied.  In an ideal linear elastic medium, a propagating mode
along the crack front with a velocity slightly less than the Rayleigh
wave velocity, is found to exist.  But the dependence of the effective
fracture toughness $\Gamma(v)$ on the crack velocity is shown to
destabilize the crack front if $\frac{d\Gamma}{dv}<0$.  Short wavelength
radiation due to weak random heterogeneities leads to this instability
at low velocities. The implications of these results for the crack dynamics
are discussed.  
\end{abstract} 
]
\narrowtext

The conditions under which solids fail by extension of pre-existing
cracks are quite well established\cite{lawn}, but the {\it dynamics}
of crack-front propagation is poorly understood.  Experiments show a
variety of types of crack surfaces---including fractal roughness over
a wide range of length scales\cite{Bou2}---and erratic dynamics of the
local crack-front velocity\cite{swinney}.  With the exception of the
uniform motion of a straight crack-front of a planar crack in an ideal
elastic solid, there has not been, until recently, much theoretical
analysis of these problems.  Indeed, even what determines in which
direction a section of crack-front will advance is not
understood\cite{crackdir,sethna,langer}, although it is clear that
this involves physics on smaller scales than the regime of continuum
elasticity theory. This has been the subject of substantial recent
attention\cite{sethna,langer}, but the effects of heterogeneities,
side branching and other effects are only just starting to be
addressed.

In this paper, we consider a seemingly much simpler problem: the
dynamics of a tensile (mode I) crack that is confined to a plane in a
weakly heterogeneous material.  In this case, the local dynamics of
the crack-front is determined by energy conservation and we can
therefore use continuum mechanics with the non-linear small scale
physics near the crack-front just determining the local fracture
toughness.  But we shall see that, nevertheless, the combined effect
of even weak heterogeneities and elastodynamics can be dramatic.

In a scalar approximation to elasticity, Perrin and Rice \cite{perrin}
found that to lowest order in a weak randomness perturbation theory,
heterogeneities in the fracture toughness led to slow---logarithmic in
time---roughening of a planar crack-front suggesting that this system
does not reach a statistical steady state. In this paper, we first
show that the dynamics of a tensile crack are very different from the
scalar approximation. In particular, in an ideal system with constant
fracture toughness, $\Gamma$, a mode I crack that is advancing with
velocity $v$, supports a {\it propagating mode} along the crack-front
with a velocity that is slightly slower than the Rayleigh wave
velocity, $c$, and depends weakly on $v$. This mode does not exist in
the scalar approximation and its presence dramatically enhances the
effects of weak heterogeneities.  Furthermore if the toughness
$\Gamma(v)$ is not constant but decreases with velocity, the
propagating mode---and hence the crack-front---becomes {\it unstable}.
A perturbative analysis of the effects of random local heterogeneities
is found to yield an effective $\Gamma(v)$ whose contributions from
short wavelength acoustic emission indeed make
$\frac{\partial\Gamma}{\partial v}<0$ at least for $v/c$ small. The
implications of these results are discussed and we speculate on
possible scenarios suggested by them.

 We study a tensile crack confined to the plane $y=0$ with the crack
open in the region $x<F(z,t)$.  The normal displacement has a
discontinuity across the crack, i.e., $u_y(y=0^+)\not= u_y(y=0^-)$,
while the normal stresses, $\sigma_{iy}(y=0^\pm)$ vanish on the crack
surfaces. Under an external load, the displacement field $u_y$ has a
$\sqrt{F(z,t)-x}$ singularity at the crack-front with an amplitude
proportional to the local mode I stress intensity factor, $K_I(z,t)$.
As long as the crack remains planar, the loading is purely mode I so
we write $K\equiv K_I$. The system is under a static load applied very
far away so that for a straight crack at rest, (i.e., $F$=constant),
$K=K^\infty$=constant.  The crack is considered to be initially at
rest (at large negative times) and move a total distance small
compared to its length and the sample dimensions and for a time less
than the sound travel times across the sample.

If the crack-front advances, $F\to F+\delta F$, an energy per unit
area of new crack $\Gamma[x=F(z,t),z]$ must be provided to the crack
front in order to fracture the solid; in an ideal quasi-equilibrium
situation this is just twice the solid-vacuum inter-facial energy
density, more generally it is the local fracture toughness that
includes the effects of small scale physics for which linear continuum
elasticity is not valid.  The fracture energy will be provided to the
crack-front by a flux of stored elastic energy per unit area of new
crack $G(z,t,\{F\})$, which in general depends on the past history of
the crack and its instantaneous local velocity $\frac{\partial
F}{\partial t}.$ The dynamics of the crack is given by
\begin{equation}
G[z,t,\{F(t'\leq t)\}]=\Gamma[x=F(z,t),z]
\label{eq:a}
\end{equation}
for all $z$ and $t$.

The available energy, $G$, has the general form
\begin{equation}
G=\Phi\left[v_\bot(z,t)\right]
G_R\left[z,t,\left\{F(t'<t)\right\}\right]
\end{equation}

where $G_R$, which is independent of $\partial F(z,t)/\partial t$), is
the available elastic energy at $(z,t)$ for a crack at rest locally
and $v_\bot=\frac{\partial F}{\partial t}/\sqrt{1+\left(\frac{\partial
F}{\partial z}\right)^2}$ is the local velocity normal to the crack
front.  For a straight crack $F=$constant,
$G_R=G^\infty=\left(K^\infty\right)^2 \left(1-\nu ^2\right) /E$, with
$E$ Young's modulus and $\nu$ Poisson's ratio \cite{freund}.  The
fraction of $G_R$ available for fracture $\Phi(v_\bot)$, decreases
linearly from unity for small $v_\bot$ and goes to zero for
$v_\bot=c$, yielding, for a straight crack with uniform toughness, a
monotonic $v(G^\infty)=\Phi^{-1}(G^\infty/\Gamma)$ for $G^\infty$
greater than the Griffith threshold, i.e., $G^\infty>\Gamma$.

In a weakly inhomogeneous material, one can hope to perturbatively
expand about a uniformly opening crack, writing
\begin{eqnarray}
F(z,t)&=&vt+f(z,t),\nonumber\\
\Gamma(x,z)&=&\Gamma_0[1+\gamma(x,z)]\nonumber\\
\text{and} \hspace{1.1415cm}G&=&G_0[1+g(z,t)].
\end{eqnarray}

The variation in the available energy for a small crack-front
distortion has the form:
\begin{equation}
g=-P\otimes f+S(f,f)+T(f,f,f)+O(f^4)
\label{eq:n5}
\end{equation}
where in Fourier space $z\to k, t\to \omega$
\begin{eqnarray}
P=|k|\left\{\right.
&&\frac{2c}{c^2-v^2}\sqrt{c^2-\sigma^2}
-\frac{a}{a^2-v^2}\sqrt{a^2-\sigma^2}\nonumber\\
&+&\frac{1}{2\pi i}\oint dW I(W,\sigma^2)
\left.\right\}
\end{eqnarray}
with
\begin{eqnarray}
I=&&\frac{2v^2W-\sigma^2 (W+v^2)}{\sqrt{W(W-\sigma^2)}(W-v^2)^2
}\nonumber\\
\times&&\ln \left[\left(2-\frac{W}{b^2}\right)^2-
4\sqrt{1-\frac{W}{a^2}} \sqrt{1-\frac{W}{b^2}}\right];
\label{eq:n7}
\end{eqnarray}
the contour integral circling the cut in the complex $W$ plane from
$W=b^2$ to $W=a^2$ with $a$ and $b$ the longitudinal and transverse
sound velocities respectively; the argument of the logarithm in
Eq.~(\ref{eq:n7}) the function whose zero is $W=c^2$;
$\sigma^2\equiv\frac{\omega^2}{k^2}+v^2$ the square of a phase
velocity in the rest frame; and $\omega\to \omega+i0$ needed to define
all cuts e.g., $\sqrt{c^2-\sigma^{2}}=-isign(
\omega)\sqrt{\sigma^{2}-c^2}$ for large $\omega$. The functionals $S$
and $T$ are bi- and tri- linear, respectively, with each consisting of
a sum of many terms that are sequences of multiplications and
convolutions ($\otimes$) involving $P(k,\omega;v)$ and functions
related to it The linear kernel, $P$, can be derived from the results
of Movchan and Willis\cite{willis} but a simpler method was developed
to obtain the non-linear terms; it will not be reproduced here.  For
large $\omega$, $P\approx-i\omega \left(-\frac{d\ln\Phi}{dv_\bot}\right)$
representing the dependence of $G$ on local velocity, while for
$\omega=0$, $P\propto|k|$ representing the long range stiffness of the
crack (as found in the scalar elasticity approximation)\cite{perrin}.
Generally, for $|\omega|<\sqrt{c^2-v^2}|k|$, the imaginary part
$P''(k,\omega)=0$ since no bulk or crack-surface modes can carry away
energy in this regime.  But, perhaps surprisingly, for a particular
$\omega<c|k|$, $P[\omega=\pm s(v)|k|]=0$ implying a propagating mode
along the crack-front with velocity $s(v)$ in the moving frame and
$\frac{s(v)}{\sqrt{c^2-v^2}}$ increasing from 0.94 for $v=0$ to 1 for
$v\to c$.  Evidence for this propagating mode is seen in recent
numerical simulations of Morrissey and Rice\cite{morrissey}. This is
in striking contrast to the scalar approximation for which
$P=\frac{c}{c^2-v^2} \sqrt{k^2(c^2-v^2)-\omega^2}$ yielding a
$1/\sqrt{t}$ decay of the maximum distortion at long times after a
localized disturbance caused by a tough region.

Damping of bulk sound modes can be taken into account by replacing
$\omega^2$ by $\omega^2/(1-i\omega\tau_d)$ in $P$ (and $S$ and $T$).
This yields a damping rate of the propagating crack-front mode
proportional to $\omega^2\tau_d$. The leading effect of weak
heterogeneities can readily be seen to yield a mean square roughness
in steady state:
\begin{eqnarray}
C(z)&\equiv&\langle\left|f(z+z',t)-f(z',t)\right|^2\rangle\nonumber\\
&=&\int_k\int_\omega
\frac{\hat\Upsilon(q
=-\frac{\omega}{v},k)(2-2\cos
kz)}{v|P(k,w)|^2}
\end{eqnarray}
with $\hat\Upsilon$ the Fourier transform of the correlations
$\langle\gamma(x,z)\gamma(x',z')\rangle$ of the fractional toughness
variations.  The propagating mode yields at large separations
$|z|\gg{cl}/{v}$
\begin{equation}
C(z)\sim\frac{|z|}{v\tau_d}\langle\gamma^2\rangle ^2
\label{eq:n9}
\end{equation}
where $l$ is a characteristic correlation length of the toughness
variations.  This is much rougher than the
$C(z)\sim\frac{cl^2}{v}\ln^2\left(|z|/c\tau_d \right)$ or
$C(z)\sim\frac{cl^2}{v}\ln\left(|z|/c\tau_d \right)$ obtained,
respectively, in the scalar or quasi-static $(P=-iw/c+|k|)$
approximations to the stress transfer.

As we shall see, the roughness of Eq.~(\ref{eq:n9}) is caused by a
straight moving crack being at the boundary of a regime of stability
so that small perturbations---which we will find can be driven by the
randomness induced fluctuations---can cause it to become much more
unstable.

We are particularly interested in the behavior on length and time
scales much longer than the scales of variations of the toughness,
i.e., $l$ and $l/v$ respectively.  On long scales, one might expect
that the crack-front would have a well defined coarse-grained position
$\tilde f$ that does not vary much, even if on small scales the local
velocities and local angles, $\partial f/\partial z$, are strongly
fluctuating.  The long wavelength and low frequency components of the
elastic energy flux into the crack, $\tilde G$ (or $\tilde g$) will
still be given by $\tilde g=-P\otimes\tilde f$.  But the non-linear
effects of the random toughness and the non-linearities in $G[\{F\}]$
will make the effective toughness $\tilde\Gamma$ include contributions
from radiation of short wavelength sound waves as well as other small
scale physics such as plasticity, etc. These cause the effective
toughness to increase and become dependent on the coarse-grained
velocity and history.  If the randomness is effectively averaged out,
on long scales the crack will be moving almost uniformly with a mean
velocity $\overline v$ determined by
\begin{equation}
\tilde\Gamma[\overline
v,\{\Gamma\}]=G^\infty \Phi\left(\overline v
\right)
\label{eq:h}
\end{equation}
with $\tilde\Gamma$ depending on the distribution and spatial
correlations of the microscopic $\Gamma(x,z)$.

Naively, it would appear that this system is stable as long as
$v(G^\infty)$ is single valued, i.e., if $\tilde
A\equiv\frac{\partial\ln\tilde \Gamma}{\partial\overline v}>\frac{d\ln
\Phi}{d\overline v}$. But, in fact, the elastodynamics implies that it
is unstable at finite wavelengths unless
$\frac{\partial\tilde\Gamma}{\partial\overline v}>0$. Linearizing in
perturbations $\tilde f$ about a putative fixed velocity steady state,
given by Eq~(\ref{eq:h}), we have for long wavelengths and low
frequencies, $-i\omega\tilde A\tilde f=-P\tilde f$.  For $\tilde A>
0$, the response function $\tilde \chi=\frac {1}{P-i\omega \tilde A}$
has no singularities in the upper half plane and thus the system is
stable.  The perturbative effects of the residual small randomness
will then result in fluctuations in $\tilde f$ that are strongly
suppressed and more like the quasi-static result, with
$C(z)\sim\frac{l^2}{v\tilde A}\ln z$ for small $\tilde A$.

With the dynamics appropriate with $\tilde A>0$, it can be shown that
non-linear terms consistent with the symmetry are irrelevant at long
scales (in contrast to the case of a simple elastic line moving
through a random potential).  Thus the long wavelength low frequency
behavior for $\tilde A>0$ is quite simple, although the dynamic
correlations will still reflect the elastodynamics unless $\tilde A$
is very large. But for $\tilde A<0,\ \tilde\chi$ will have poles in
the upper half plane corresponding to {\it unstable} propagating modes
with $\omega\approx\pm s(v)|k|-i \eta(v)|k|$ with $\eta\propto\tilde
Ac^2<0$.

Since we have begun with a more microscopic description of the system,
it should be possible to compute the large scale response function
perturbatively for weak heterogeneities and thereby find the sign of
$\tilde A$ for the case in which the microscopic $\Gamma(x,z)$ is {\it
not} velocity dependent. It is somewhat simpler to compute a slightly
different quantity: $\chi(k,\omega)\equiv\frac{\delta f(k,\omega)}
{\delta \epsilon(k,\omega)}$ which is the response to extra energy per
unit area added at the crack-front (or roughly equivalently a decrease
in the toughness); this will yield the behavior of perturbations and
hence also the randomness induced fluctuations.

The unperturbed response function is $\chi=\chi_o=\frac{1}{P} $. Since
the interesting regime is near where $P$ vanishes, a direct expansion
of $\chi$ in powers of $\Upsilon$ is not useful.  Thus instead, we
expand
\begin{equation}
\chi^{-1}\equiv P-\Sigma
\label{eq:j}
\end{equation}
with $\Sigma$ (the analog of a ``self-energy'') expected to be well
behaved at small $k$ and $\omega$; we will thus focus on
this. Conventional Feynman diagrammatic techniques can be used via a
Martin-Siggia-Rose formulation of the dynamics, analogous to
treatments of, e.g., charge density wave dynamics\cite{cdw}.  The main
complication here is the form of the non-linearities. In addition to
the non-linearities from expanding $\Gamma[x=vt+f(z,t)]$ in powers of
$f$, there are also the non-linear terms in $G[\{F\}]$,
Eq.~(\ref{eq:n5}), arising from the moving boundary problem of an
irregular crack-front.  To compute even the leading correction to
$\chi^{-1}$, both the second and third order non-linearities in $G$
are needed.  But for velocities $v\ll{c}$, only the second order
parts, $S$, are needed and there are some simplifications; we will
focus on this regime here.

From the structure of the perturbative expansion, it can be seen that
$\Sigma(k,\omega)$ vanishes for $k=\omega=0$, and for small $k$ and
$\omega$ has the form
\begin{equation}
\Sigma(k,\omega)\approx i\omega
A+BP(k,\omega)+O(\omega^2, k^2, P^2, P_2,
i\omega^3\dots)
\label{eq:k}
\end{equation}
 with $P_2=\int^\omega
d\omega'\frac{\partial P(\omega')} {\partial
v}.$
From the above discussion, it should be clear that the coefficient $A$
will be the most important, with $\tilde A\approx A[1+O(B)]$. We are
thus interested primarily in the weak randomness expansion of $A$.

For low velocities, inverse powers of $\overline v$ are associated
with non-linearities from the random toughness and thus would be
expected to dominate, with the natural expansion parameter being
$\langle\gamma^2\rangle c/\overline v$\cite{cdw}. But the leading low
velocity contribution to the expansion of $A$ is found to vanish
exactly.  The dominant term thus involves both the non-linearities
from the random toughness and those from the changes in crack shape.
Evaluating these terms yields

\begin{equation}
A\approx-\frac{1}{2}\int^\infty_{-\infty} \frac{dq|q|}{2\pi}
\Upsilon(q,k=0) \int^\infty_{-\infty}
\frac{d\omega|k|}{2\pi\omega^{2}}\frac
{\left[Im\,P(k,\omega)\right]^2} {\left|P(k,\omega)\right|^2}
\label{eq:nG}
\end{equation}
which is independent of v.  Corrections to this arise primarily from
the randomness induced non-linearities with $\langle\gamma^2\rangle
^2/v^2$ terms expected at next order.  Thus the expansion should be
good if $\langle\gamma^2\rangle{\ll}v^2/c^2$ (rather than the naive
$\langle\gamma^2\rangle{\ll}v/c$).

  At low velocities, we thus conclude that the crack-front is unstable
at long wavelengths, $\lambda>\lambda_c$.  The $i\omega^3$ term in
$\sum$ and the sound wave damping stabilize the system for shorter
wavelengths, $\lambda<\lambda_c$, with $ \lambda_c\sim max
\left[\frac{c\tau_d} {\langle\gamma^2\rangle},\ l\left(
\frac{c}{v}\right)^{3/2}\right]; $ note that the second cutoff
corresponds to modes with period much longer than the time,
$\frac{l}{v}$, for the crack to advance a distance $l$.  The most
rapidly growing modes will have $\lambda\sim 2\lambda_c$.

At higher velocities, there are contributions of both signs to the
order $\gamma^2$ terms in $\Sigma$ and the sign of $A$ depends on more
details of the correlations, $\Upsilon(p,k)$, but at least for simple
forms of the correlations, $A$ can become positive for high velocities
and the crack-front will be stable.\cite{scalar}

We now consider various possible scenarios for mode I planar cracks.
The simplest scenario, suggested by the perturbative analysis, is that
there will be a regime of unstable velocities probably extending down
to zero velocity---although the perturbative analysis certainly fails
there.  Associated with this, one might expect hysteresis in the
macroscopic velocity $\overline v$ versus the load $K^\infty$ (or
$G^\infty$) with a stationary crack jumping to a finite velocity as
the load is increased, with a statistically steady state moving phase
that is only logarithmically rough---corresponding to $\tilde
A(\overline v)>0$.  On decreasing the load, the crack would not stop
until it reaches some lower critical load at which $\tilde A\to 0$
whereupon it would stop suddenly. Alternatively, for macroscopic
systems, the hysteresis might disappear leaving a single-valued but
discontinuous $\overline v(G^\infty)$ and a strongly fluctuating crack
for $G^\infty$ just above the critical load
$G^\infty_c\approx\overline{\Gamma}[1+ O(\langle\gamma^2\rangle)]$. The
scenario with the stationary crack jumping to a finite velocity when
the loading is increased, is consistent with our numerical results on a
simplified model.\cite{numerics}

Another possibility is that strongly non-linear fluctuations in the
local velocity driven by the variable toughness might drive $\tilde A$
positive and stabilize the crack dynamics even for low velocities. But
our understanding of the behavior near $G^\infty_c$ in the
quasi-static approximation suggests that this is
unlikely\cite{scenario2}. Perhaps the most interesting possibility is
that for some range of velocities, a non-trivial, non-linear
statistical steady state might occur that is much rougher than
logarithmic and perhaps characterized by propagating modes along the
crack. This would be roughly analogous to the behavior of the
$KPZ$\cite{kardar} equation in dimension $d>2$ for which both a
non-trivial strongly fluctuating phase and a simple weakly fluctuating
phase exist.  Combinations of these possibilities could also occur. 
The behavior should depend on aspects of the correlations of the
random toughness as well as its magnitude and on velocity dependent
toughness that arises from other physics such as plasticity,
dislocation emission, thermal creep, and small scale side branching.

 Note that even in an ideal system with no toughness variations,
thermal fluctuations can give rise to effects similar to those
discussed here as the elastic energy needed to break a bond is less if
the bond is already stretched due to thermal fluctuations.  For $v$ a
substantial fraction of $c$, the growth rate of the fluctuation driven
instability will be of order $\frac{1}{\tau}\delta^4$ with $\delta$ of
order the ratio of the temperature to the bond breaking energies, and
$\tau$ a microscopic time.  In practice, the effects of the atomic
scale physics on the effective $\tilde\Gamma(v)$ will probably
dominate over thermal fluctuation effects.

One type of correlation in the toughness that is important to consider
is {\it periodic} inhomogeneities---a crude approximation to some of
the effects of a crystal lattice.  In this case, weakly variable
toughness can again be shown to lead to a perturbatively unstable
crack-front at low velocities. But the short wavelength acoustic
radiation that leads to the velocity dependent effective toughness
arises from high order harmonics---of order $\frac{c}{\overline v}$
---of the periodicity in the $x$-direction; thus at low velocities the
leading perturbative effect can be strongly suppressed.  This
instability caused by non-steady motion in a periodic system would
appear to be related to that found by Marder and Gross in a
one-dimensional ``ball and spring'' model of a crack\cite{marder}.

We have shown here that both random and periodic inhomogeneities in
the toughness can dynamically destabilize the front of a tensile crack
moving in a plane and in general such cracks will be unstable to
finite wavelength disturbances if their effective toughness decreases
with velocity.  Understanding the resulting dynamic states of the
system as well as the more challenging problem of out-of-plane crack
motion, we must leave for future work.

We would like to thank J.~R.~Rice and J.~Morrissey for useful
discussions and communicating their numerical results. This work has
been supported in part by the NSF via DMR-9106237, 9630064 and Harvard
University's MRSEC.
 

\begin{references}
\bibitem{scalar}We note that in the scalar model rather different
physics occurs: the short wavelength radiations lead to
$\frac{d\Gamma}{dv}<0$ resulting in a supersonic mode at low
velocities . This mode is generically weakly unstable, resulting again
in the instability of the moving crack-front. At higher velocities,
depending on the microscopic details of the system,
$\frac{d\Gamma}{dv}$  can change sign resulting in stable crack-fronts
that are only logarithmically rough
\end{references}
\end{document}